# Low-Afterglow, High-Refractive-Index Liquid Scintillators for Fast-Neutron Spectrometry and Imaging Applications

Ronald Lauck, Michal Brandis, Benjamin Bromberger, Volker Dangendorf, Mark B. Goldberg, Ilan Mor, Kai Tittelmeier and David Vartsky

*Abstract*—For ion and neutron spectrometry and imaging applications at a high intensity pulsed laser facility, fast liquid scintillators with very low afterglow are required. Furthermore, neutron imaging with fiber (or liquid-core) capillary arrays calls for scintillation materials with high refractive index. To this end, we have examined various combinations of established mixtures of fluors and solvents, that were enriched alternatively with nitrogen or oxygen. Dissolved molecular oxygen is known to be a highly effective quenching agent, that efficiently suppresses the population of the triplet states in the fluor, which are primarily responsible for the afterglow. For measuring the glow curves of scintillators, we have employed the time-correlated single photon counting (TCSPC) technique, characterized by high dynamic range of several orders of magnitude in light intensity. In this paper we outline the application for the fast scintillators, briefly present the scintillation mechanism in liquids, describe our specific TCSPC method and discuss the results.

*Index Terms*—Liquid scintillator, high refractive index liquid, light decay characteristic, afterglow, single photon counting.

## I. INTRODUCTION

HIGH intensity lasers have been proposed in recent years as radiation sources for various scientific, medical and industrial applications. Present-day high-power laser systems generate relativistic electrons, X- and γ-rays, neutrons and energetic ions by the interaction of ultra-short light pulses of petawatt power density with matter [see e.g. 1 – 4].

In this context, we have launched a programme to perform the spectrometry of the associated fast (i.e. 1 MeV – several tens of MeV) neutrons produced, in the light of possible future neutron imaging applications. These neutrons can either be produced directly in a plasma (e.g., by the d-d fusion reaction) or preferably, by the ions accelerated from a laser-plasma (protons, deuterons) using appropriate secondary targets (Li or Be). Such experiments will be performed at the POLARIS high intensity laser facility of Jena University [5].

For neutron spectrometry using pulsed ion beams produced by femtosecond (fs) laser pulses, Time-Of-Flight (TOF) spectrometry is the appropriate method, employing a well characterized fast organic scintillation detector [4]. A severe difficulty in such experiments is the intense γ-burst, that precedes the ions and neutrons in TOF. Suitable lead shielding reduces the gamma signal. Furthermore, with adequate gating techniques of the phototube, one can further suppress the response of the scintillator to the gamma flash. However, the huge intensity of this gamma burst still tends to mask the much weaker neutron signals, because the afterglow of a fast organic scintillator continues to dominate the signal in the TOF region where the neutrons are expected. Hence, an organic scintillator with fast light decay characteristics (main decay time < 2 ns for sufficiently good TOF resolution) and very low afterglow is required for these measurements. In a typical experiment of this type, with a flight distance of ~2 meters, the first neutrons or ions will reach the spectrometer or imaging system ~50 ns after the γ-burst. Within this time interval, the γ-induced scintillator light signal needs to decay by 4 – 5 orders of magnitude of its peak intensity, in order to render the relatively weak neutron or ion signals detectable. Fig. 1 shows the γ-signal measured with a very fast plastic scintillator (Pilot U of formerly Nuclear Enterprise, today EJ-228 or BC-418 [6]) and a fast Photonis XP2020 photomultiplier tube (PMT) after being exposed to the γ-ray burst from a thick deuterated polyethylene target.

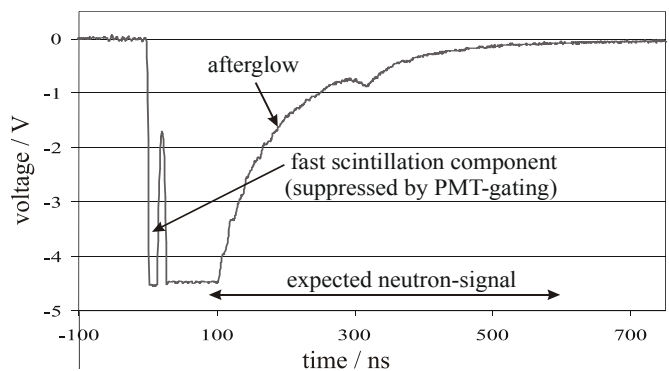

Fig. 1. Measured PMT anode signal from a "Pilot U" fast plastic scintillator after being exposed to a Laser induced the γ-ray burst. A neutron event would show up in the indicated region as a few ns wide pulse with an amplitude of a few-hundred mV.

The target was hit by the pulse of the Jena Terawatt laser "JETI". Though the main pulse during the first 20 ns is largely suppressed by dynode gating (though still visible in the remaining peak at time $t = 0$ in the diagram), the afterglow at $t > 20$ ns, when the gate opens, still saturates the PMT during the first 100 ns and completely masks any possible neutron event up to ca 250 ns after the γ-burst has hit the scintillator. The small bump at 300 – 350 ns is due to ion feedback in the PMT. For practical reasons we have focused our efforts exclusively on the development of low afterglow **liquid** scintillators.

As mentioned above, the future application would be a neutron imaging system, e.g., the type of device described in [7]. Here, rather thick neutron converter plates (20 – 30 mm thickness) are required to achieve substantial detection efficiencies ( > 10 % is mandatory). In the detector described in [7] the use of scintillating plastic fiber or liquid scintillator filled capillary fiber arrays for optimal position resolution and high detection efficiency is demanded. To optimize the light output of such a capillary fiber array, liquids with as high a refractive index as possible are required. Therefore, our search has covered not only fast, low-afterglow liquids but also those with high refractive index.

## II. THEORY OF LIQUID SCINTILLATORS

The basic components of an organic liquid scintillator are a solvent (e.g., p-Xylene) and one or more fluorescent solutes (e.g. PPO, di-methyl-POPOP, bisMSB). In organic liquid scintillators both, solvent and solute, are aromatic hydrocarbons which are characterized by benzenoid and heterocyclic ring structures [8, 9]. Each atom in the ring structure is bonded to its neighbor by localized σ-electrons. In addition, there exists a collective bond of delocalized π-electrons. This π-bond is far less tightly bound than the σ-bond. However, it is of crucial importance for the scintillation properties that the emission spectra of the π-orbitals are in the visible or near ultraviolet region.

The primary origin of scintillations in organic liquid scintillators is the excitation of these π-electrons following an interaction with energetic particles. The processes which determine the de-excitation (and thereby, the fluorescence) are manifold and highly complex [8, 9]. For this reason only the fundamental interactions and processes are discussed here.

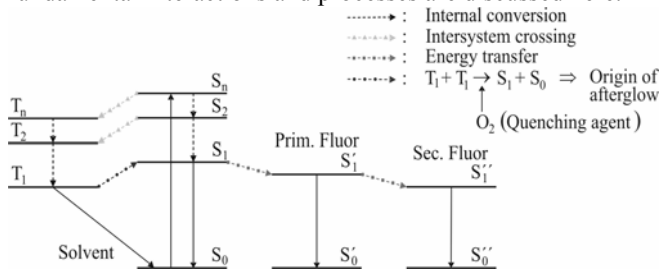

Fig. 2. Level scheme of a ternary mixture liquid scintillator, illustrating the scintillation mechanism. Most of the excitation energy is originally deposited in the solvent, in which rather long-lived states are excited. The energy is first transferred to a primary and thence to a secondary fluor, that progressively shift the wavelength of the emitted light to the visible.

The collective spin of the π-orbitals could be 0 or 1. Thus two series of excited states exist, the singlet states $S_n$ and the triplet states $T_n$, as schematically shown in Figure 2. Spin-forbidden radiative transitions between triplet states and singlet states, such as $T_1 \rightarrow S_0$, are very rare. The corresponding light emission thus exhibits decay times up to milliseconds and is called phosphorescence. In contrast, the spin-allowed scintillation emission which originates from $S_n$ occurs within a few ns and is responsible for the fast scintillation component.

The fluorescent solutes introduced at typical concentrations of 0.1 – 1%, convert the scintillation wavelengths to visible or near ultraviolet light by radiationless energy transfer, that occurs in collisions between the molecules. Due to these fluors, the scintillator is rendered transparent to its own light and the effective life time of the $S_1 \rightarrow S_0$ transitions is shortened, to a certain extent.

The scintillation emission which originates from triplet-triplet-interactions and decays to $S_1$, such as $T_1 + T_1 \rightarrow S_1 + S_0$ occurs within a few μs and is responsible for the delayed component. This component is independent of the presence of wavelength-shifting fluors in the scintillation mixture. In certain cases, a secondary solvent (or solute) is introduced at typical concentrations of ~10%, in order to enhance it. The relative intensities of the fast and slow components depend on the specific ionization density of the charged particle (commonly denoted *LET* or *dE/dx*) and constitute the basis for Pulse Shape Discrimination (PSD) [9, and references therein] – a technique commonly used to distinguish between different types of radiation.

Liquid scintillators containing dissolved molecular oxygen are characterized by a reduction in scintillation efficiency and a suppression of the slow component [9, 10]. As an electron acceptor, oxygen can therefore compete with radiation processes that yield π-electronic excited species. Via the $T_1 + {}^3O_{20} \rightarrow S_0 + {}^1O_{21}$ process, oxygen also quenches the excited triplet states that are responsible for the slow scintillation component.

## III. EXPERIMENTAL SETUP

The light decay characteristics of the scintillators were investigated by means of the Time-Correlated Single Photon Counting technique (TCSPC) [11]. With the setup schematically shown in Fig. 3 we measured light intensities over 5 orders of magnitude. This high dynamic range is the principal advantage of this method. A further advantage is the independence of measured values from variations of single photon pulse height and gain. A disadvantage is the long measurement time of several days per scintillator.

In our TCSPC setup we irradiated the liquid scintillator in question, as well as a fast, highly-efficient $BaF_2$ scintillator, with the two time-correlated 511 keV γ-rays emitted by a $^{22}$Na source. The $BaF_2$ was coupled to its PMT with the highest achievable light collection efficiency - in particular it was important to detect the fast UV component of the $BaF_2$ luminescence to achieve a fast, sub-ns trigger signal. In contrast, the light coupling of the liquid scintillator to its PMT was deliberately degraded by an aperture to such a low level that, per 10 absorbed gammas, only ~1 photoelectron was produced by the photocathode of the PMT. Thus it was ensured that ~90% of all events detected by this PMT originated from a single scintillation photon.

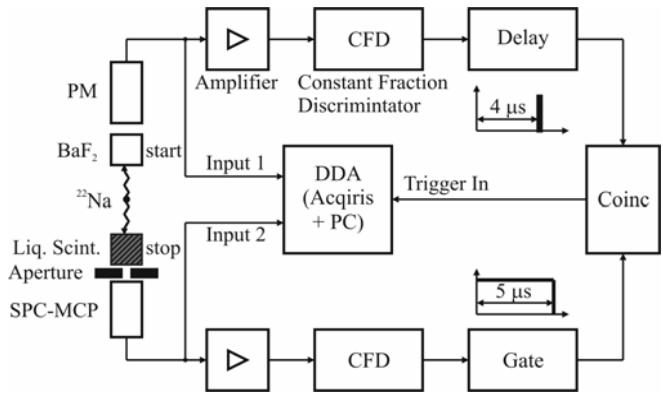

Fig. 3. Schematics of the experimental setup for measuring glow curves of liquid scintillators. The method is known as Time-Correlated Single Photon Counting (TCSPC) and permits measuring the light decay of the scintillation down to intensity levels 4 to 5 orders of magnitude below peak luminescence. CFD is a constant fraction discriminator, Coinc is a coincidence logic unit.

Special care was required in the choice of the PMT for the liquid scintillator. Regular PMTs with a long electron flight path from photocathode to the $1^{st}$ dynode are prone to afterpulsing from back-drifting ions. These ions are produced by energetic electron collisions with residual gas and, due to the low average velocity of the ions, are delayed by several hundred nanoseconds. For the fastest of the scintillators measured in this programme, these afterpulses are more intense than the delayed scintillation light.

To minimize the contribution of ion feedback we chose a proximity-focused MCP image intensifier (MCP-PMT) instead of a regular PMT. In such a MCP-PMT the gap between photocathode and MCP is as small as 0.1 mm. A chevron arrangement of the 2 MCPs ensured negligible ion feedback from the anode side of the tube. In practice, some ion feedback signals from electron collisions in the photocathode gap or inside the $1^{st}$ MCP channels are still visible. However, since their delay is of the order of 20 – 40 ns, they are too small to be visible in the delayed light emission, even for the fastest liquid scintillators.

The PMT-anode signal of the $BaF_2$ trigger detector and the time-correlated single photoelectron event in the liquid scintillator were recorded by a digital data acquisition system (DDA), based on a dual channel Acquiris DC240 fast digitizer [12]. Both signals were sampled at a rate of 2 G-Samples/s during an observation interval of 5μs. Coincident signals between $BaF_2$ and liquid scintillator were selected by a fast logic unit and provided a trigger for the Acquiris DDA. The high dynamic range (of order $10^5$) in light intensity calls for an extremely small random coincidence event rate (i.e. uncorrelated gammas detected during the observation time window of 5 μs). This implies that the single rates in the detectors need to be rather low which, in turn, means that the measuring times per scintillator sample extends over 2 – 4 days.

Classical TCSPC methods take the time differences of the first registered signals of each detector [11]. Compared to this, the digital data acquisition with its capability of complete and multi-hit signal recording offers several advantages: By defining suitable software filters in the analysis of the single-photon pulses unwanted events, such as random coincidences, remnant afterpulses from the PMT, or electronic pick-up signals, can be suppressed. Thus, the measurement accuracy is significantly improved, particularly with respect to the dynamic range of the light intensity in the delayed component. Moreover, the effect of software-corrected glow curves, compared to those obtained using unfiltered "raw" coincidence events, resulted in an improvement of the dynamic range by a factor of 10! Using this filter technique, we can indeed track the level of the afterglow over 5 orders of magnitude in light intensity.

The total light yield of the liquid scintillators was measured by comparing the position of the Compton edge of the 511 keV and 1275 keV γ-spectra produced by a $^{22}$Na source. For this measurement the scintillator cell was coupled with good optical contact to a Photonis XP2020 PMT. The anode pulse was sampled by the DDA system and integrated by software. The pulse height was corrected for the wave-length dependent sensitivity of the PMT photocathode using the sensitivity curves of the manufacturer. No attempt was made to measure the absolute light output – all values relate to the nominal light output of a NE213 scintillator. We also compared only those scintillator liquids that could be filled into the same standard cell – to avoid errors due to different geometries and light collection properties. For this reason the solid scintillators (like e.g. the Pilot U) are not included in this comparison.

IV. RESULTS AND DISCUSSION

In this report we present some typical results on several classes of scintillators which have been investigated so far. A more complete presentation, which includes all scintillation mixtures we investigated and their relative magnitudes and time constants of their slow scintillation components will be published elsewhere [13]. Fig. 4 shows the glow curves of 7 different scintillators and Table 1 shows a comparison of their relative light output and other data relevant to neutron detection and imaging. As mentioned above, Pilot U is a very fast plastic scintillator made by the former Nuclear Enterprises Ltd. "for very fast timing applications". Clearly, this scintillator, fast as its primary light component might be (it is quoted to be 1.4 ns [6]), has a very intense afterglow in the few-microsecond range. It is therefore completely unsuitable for the application in view. As a matter of fact, these "ultrafast" plastics were optimized for the shortest possible **fast** light component, without paying attention to the afterglow, which is apparently of no importance in their conventional applications. By comparison, we have also measured the slow component of our "standard work horse" in fast neutron detection, the NE213 liquid scintillator (today available as EJ-301 from Eljen or BC-501A from St. Gobain). This liquid was developed to discriminate between neutrons and gammas by determining the relative magnitude of fast to slow fluorescence; therefore, a significant contribution of the slow component is essential in the scintillation output. It can be seen that the slow component of NE213 is comparable to that of Pilot U. Similar intense afterglow was observed for nitrogen-doped Methylnaphthalene based scintillators. Methylnaphthalene as primary solvent of a liquid scintillator is of interest due to its high refractive index (see Table 1), which renders it a promising candidate for a capillary-fiber-based neutron-converter array [14]. Fig. 4 shows that a nitrogen-doped cocktail of this type has a similar afterglow to NE213 but, in addition, has an intense intermediate scintillation component in the time range which is of greatest interest to us.

Based on a suggestion by F.D. Brooks [10] we have focused our work on binary and ternary mixtures with p-Xylene and toluene as solvents. PPO (2, 5 *Diphenyloxazole*) as primary fluor and POPOP *(p-bis[2-(5-Phenyloxazolyl)]-benzene)* or bis-MSB (*bis(o-methylstyryl)benzene*) as secondary fluors. Binary mixtures with PPO as fluor emit in the UV (emission maximum ca 350 nm). Adding the secondary fluor shifts the emitted wavelength to the 420 nm region. Comparison between binary and ternary cocktails did not show much difference in the afterglow whereas, the total light yield of the ternary mixtures was significantly higher. However, this may be due to the enhanced reflectivity of the commercial reflector paint used to coat the scintillation cell [15] in these measurements. For a future imaging system it is also advantageous to have a visible light scintillator because standard optical components can be used. Ternary cocktails of the type tested here are therefore preferred over binary ones.

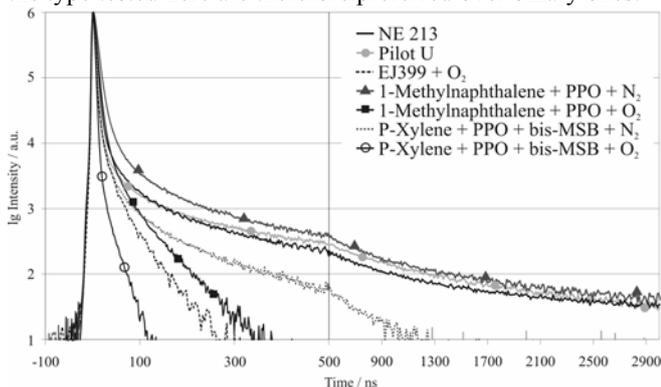

Fig. 4. Glow curves of various scintillators. Note the change in the horizontal scale at 500 ns.

The real breakthrough in the efforts to suppress the long scintillation component was the use of oxygen-saturated cocktails (suggested by F.D. Brooks [10]). As can be seen in Fig. 4, the long component of the p-Xylene based cocktails, already relatively small when saturated with inert gases, such as $N_2$ or Ar, could be significantly further suppressed when dissolving $O_2$ into the cocktail. Despite the fact that, with oxygen, the overall light output is reduced by a factor of more than 2 (compared to the scintillators saturated with inert gas, the light output of the optimal mixture is still comparable to that of NE213). Based on this appreciably high light yield and the negligible afterglow, p-Xylene, PPO and bis-MSB saturated with $O_2$ is the scintillator that is best-adapted to our requirements for the neutron spectrometer in the high-power laser application. The use of Toluene as solvent yielded the same fast response and low afterglow as the p-Xylene-based cocktail. Employing these scintillators in the imaging application with capillary fibers is more problematic because the refractive index is very similar to that of the Pyrex glass matrix, which serves as cladding for the liquid core.

In Fig. 4 and Tab. 1 it is also evident that oxygen-quenched Methylnaphthalene based scintillators are not an option due to their low light output and high afterglow. Those liquids are also unstable and become yellowish with time and their light output decays further when enriched with oxygen. We have also tested another commercial high refractive index liquid scintillator: EJ-399-05c1 from Eljen Tech.[16]. We measured $N_2$ and $O_2$ saturated cocktails but found that also the oxygen saturated mixture is inferior to our p-Xylene-based ternary mixture, both, in light yield and in the amount of afterglow.

| Scintillator | relative light yield (%) | H/C | density (g/cm$^3$) | refr. index | $\lambda_{emission}$ (nm) |
|---|---|---|---|---|---|
| NE213 | 100 | 1.2 | 0.87 | 1.53 | 420 |
| EJ399 | 77 | 0.91 | 0.99 | 1.56 | 420 |
| 1-MN+PPO+$N_2$ | 71 | 1.2 | 1.02 | 1.62 | 420 |
| 1-MN+PPO+$O_2$ | 29 | 1.2 | 1.02 | 1.62 | 420 |
| p-Xylene+PPO+$N_2$ | 56 | 1.2 | 0.86 | 1.52* | 350 |
| p-Xylene+PPO+$O_2$ | 32 | 1.2 | 0.86 | 1.52* | 350 |
| p-Xylene+bisMSB+$N_2$ | 219 | 1.2 | 0.86 | 1.52 | 420 |
| p-Xylene+bisMSB+$O_2$ | 103 | 1.2 | 0.86 | 1.52 | 420 |

Table 1. Selection of scintillating liquids tested in this work. The relative light yield was measured always in the same scintillation cell and refers to the yield of NE213. *The refractive index refers to light at 420 nm wavelength

## V. SUMMARY

In this paper we have presented an improved experimental setup of a Time-Correlated Single Photon Counting (TCSPC) system for measuring glow curves of scintillators, using a digital data acquisition system. This DDA system enables efficient filtering and background reduction of coincidence spectra by means of software. This technique enabled us to measure glow curves of scintillators over 5 orders of magnitude in dynamic range. Using this system we investigated various liquid scintillator cocktails and fast plastic scintillators. The goal was to optimize their response for use in high-power Laser applications, where neutrons and ions have to be detected in the presence of an intense γ-flash. When a ternary mixture of p-Xylene as solvent and PPO and bis-MSB as fluors is saturated with molecular oxygen, we have obtained a relatively bright liquid scintillator with negligible afterglow, well suited for the above application. High optical index liquid scintillators, which are required in capillary fiber arrays for fast neutron imaging, were also investigated. However, those tested were not satisfactory in their performance regarding brightness and their relatively high intensity of the slow scintillation component.

Acknowledgement:

We wish to thank F.D. Brooks from University of Cape Town for sharing his rich knowledge in the mechanisms of liquid scintillation and the fruitful discussions.